\documentclass[aps,twocolumn,groupedaddress]{revtex4-1} %fleqn
\usepackage{graphicx}
\usepackage{amsmath}
\usepackage{amssymb}
\usepackage{braket}
\usepackage{import}
\usepackage{siunitx}
\usepackage{color}
\usepackage[normalem]{ulem}
\usepackage{multirow}
\usepackage{booktabs}
\usepackage{tikz}
\usetikzlibrary{calc,patterns,angles,quotes}
\usepackage{tikz-3dplot}
\usepackage{pgfplots,pgfplotstable}
\usepackage{comment}
\usepackage{cleveref}
\crefname{equation}{Eq.}{Eqs.}  
\Crefname{equation}{Equation}{Equations}	
\crefname{figure}{Fig.}{Figs.}
\Crefname{figure}{Figure}{Figures}
\crefname{chapter}{Ch.}{Chs.}
\Crefname{chapter}{Chapter}{Chapters}
\crefname{section}{Sec.}{Secs.}
\Crefname{section}{Section}{Sections}
\crefname{appendix}{App.}{App.}
\Crefname{appendix}{Appendix}{Appendices}	
\crefname{algorithm}{Alg.}{Algs.}
\Crefname{algorithm}{Algorithm}{Algorithm}
\crefname{table}{Table}{Tables}
\Crefname{table}{Table}{Tables}

\let\originalleft\left
\let\originalright\right
\renewcommand{\left}{\mathopen{}\mathclose\bgroup\originalleft}
\renewcommand{\right}{\aftergroup\egroup\originalright}
\begin{document}
	\cmidrulewidth=.03em
	
\title{Anisotropic Stark shift, field-induced dissociation, and electroabsorption of excitons in phosphorene}

\author{H{\o}gni C. Kamban}
\email{hck@mp.aau.dk}
\affiliation{Department of Materials and Production, Aalborg University, DK-9220 Aalborg \O st, Denmark}
\affiliation{Center for Nanostructured Graphene (CNG), DK-9220 Aalborg \O st, Denmark}
\author{Thomas G. Pedersen}
\affiliation{Department of Materials and Production, Aalborg University, DK-9220 Aalborg \O st, Denmark}
\affiliation{Center for Nanostructured Graphene (CNG), DK-9220 Aalborg \O st, Denmark}
\author{Nuno M. R. Peres}
\affiliation{Department and Centre of Physics, and QuantaLab,University of Minho, Campus of Gualtar, 4710-057, Braga, Portugal}
\affiliation{International Iberian Nanotechnology Laboratory (INL),Av.  Mestre José Veiga, 4715-330, Braga, Portugal}

\date{\today}

\begin{abstract}
	We compute binding energies, Stark shifts, electric-field-induced dissociation rates, and the Franz-Keldysh effect for excitons in phosphorene in various dielectric surroundings. All three effects show a pronounced dependence on the direction of the in-plane electric field, with the dissociation rates in particular decreasing by several orders of magnitude upon rotating the electric field from the armchair to the zigzag axis. To better understand the numerical dissociation rates, we derive an analytical approximation to the anisotropic rates induced by weak electric fields, thereby generalizing the previously obtained result for isotropic two-dimensional semiconductors. This approximation is shown to be valid in the weak-field limit by comparing it to the exact rates. The anisotropy is also apparent in the large difference between armchair and zigzag components of the exciton polarizability tensor, which we compute for the five lowest lying states. As expected, we also find much more pronounced Stark shifts in either the armchair or zigzag direction, depending on the symmetry of the state in question. Finally, an isotropic interaction potential is shown to be an excellent approximation to a more accurate anisotropic interaction derived from the Poisson equation, confirming that the anisotropy of phosphorene is largely due to the direction dependence of the effective masses. 
\end{abstract}

\maketitle

\section{Introduction}

With the experimental discovery of graphene in 2004 \cite{Novoselov2004}, interest in two-dimensional (2D) materials increased enormously. Just a few years later, the successful exfoliation of monolayer MoS$_2$ \cite{Mak2010} produced the first atomically thin direct band gap semiconductor. One of the newest members of the 2D semiconductor family is monolayer black phosphorus (BP), referred to here as phosphorene \cite{Li2014,Koenig2014,Liu2014,Castellanos-Gomez2014,Xia2014}. It has seen a remarkable rate of growth in research interest, even more so than graphene \cite{Castellanos-Gomez2015}. Unlike transition metal dichalcogenides (TMDs), where the band gap is direct only in their monolayer form \cite{Chhowalla2013}, BP is a direct band gap semiconductor regardless of the number of layers \cite{Keyes1953,Asahina1984,Morita1986,Rudenko2014}. The magnitude of the gap evolves from around $0.3$ eV in its bulk form to around $2$ eV in monolayers \cite{Castellanos-Gomez2014,Qiao2014,Tran2014,Liu2014,Wang2015}. The extreme tunability of the band gap, as well as its highly anisotropic nature, make phosphorene an exceptionally interesting material for both practical applications and theoretical investigation. For instance, the tunable gap makes phosphorene a promising material for converting solar energy to chemical energy \cite{Hu2016}. The anisotropy of phosphorene shows up in almost all of its physical properties, such as its electrical \cite{Fei2014,Xu2015}, thermal \cite{Ong2014,Jain2015}, and mechanical \cite{Wei2014} properties. Particular examples of highly anisotropic quantities in phosphorene are the conductivity \cite{Xia2014b}, optical absorption \cite{Xia2014}, and photoluminescence \cite{Wang2015}. 

Very strong absorption peaks have been observed in monolayer and bilayer phosphorene \cite{Liu2014,Zhang2014}, due to strongly bound excitons \cite{Tran2014,Rodin2014,Hunt2018,Henriques2020}. How such strongly bound excitons interact with external electric fields is an interesting area of study, particularly in phosphorene as the direction of the in-plane field will matter. An applied electric field pulls electrons and holes in opposite directions, which causes a shift in the exciton energy and may even lead to dissociation of the exciton. Collectively, these effects have been studied intensely in carbon nanotubes \cite{Perebeinos2007,Mohite2008,Kamban2020_PML}, as well as monolayer \cite{Pedersen2016ExcitonStark,Haastrup2016,Scharf2016,Massicotte2018,Cavalcante2018,Kamban2019}, bilayer \cite{Kamban2020}, and multilayer TMDs \cite{Pedersen2016ExcitonIonization}. They have been studied to a lesser degree in phosphorene \cite{Chaves2015,Cavalcante2018}, where focus has been on the energy shift rather than field-induced exciton dissociation. One of the motivations for applying external electric fields to low-dimensional semiconductors from a device perspective is to induce exciton dissociation, and thereby improve photocurrent generation in, e.g., solar cells and photodetectors. The exciton Stark effect is also promising as a means of manipulating the properties of semiconductors. For instance, the shift in exciton energy and possibility of dissociation caused by an applied field shifts and broadens the optical absorption peaks. This is known as the Franz-Keldysh effect \cite{Franz1958,Keldysh1958}, and was studied for monolayer TMDs in Ref. \cite{Pedersen2016ExcitonStark}. In TMDs, both Stark and Franz-Keldysh effects are independent of the direction of an in-plane electric field. In contrast, the highly anisotropic nature of phosphorene leads to a Stark effect that is strongly dependent on the direction of the field \cite{Chaves2015,Cavalcante2018}, and this effect should also be visible in the Franz-Keldysh effect.

In the present paper, we study the exciton Stark and Franz-Keldysh effects in phosphorene. The paper is structured as follows. In \cref{sec;excitons}, we introduce the model used and show that using an isotropic interaction potential between the electron and hole is an excellent approximation to a more accurate anisotropic interaction. For phosphorene, the majority of the anisotropy therefore comes from the direction dependent effective masses. In this section, we also compute the energies of the five lowest lying exciton states in phosphorene in three different dielectric surroundings and discuss their symmetries. In \cref{sec:Stark}, the focus is on the exciton Stark effect. Here we observe Stark shifts and exciton dissociation rates that are strongly dependent on the direction of the field. To better understand these effects, we compare the Stark shifts and dissociation rates to analytical approximations derived from perturbation theory and weak-field asymptotic theory (WFAT), respectively. In \cref{sec:FK}, we turn to the Franz-Keldysh effect, which also exhibits pronounced direction dependence. Finally, the results are concluded upon in \cref{sec:conc}. The text is supplemented by two appendices. In \cref{app:poisson}, the exciton interaction potential in an anisotropic semiconductor is discussed, and in \cref{app:WFAT}, an anisotropic weak-field approximation for the exciton dissociation rate is derived.

\section{Excitons in phosphorene}\label{sec;excitons}
\begin{figure}[t]
	\includegraphics[width=1\columnwidth]{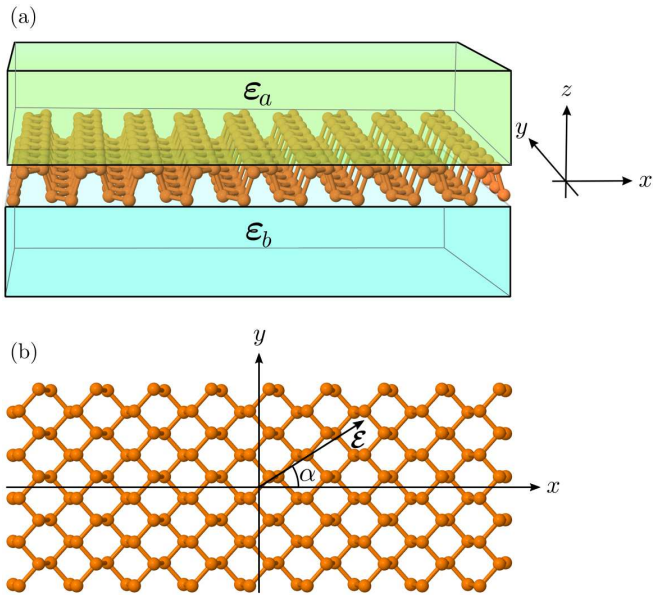}
	\caption{Sketch of the geometry. (a) side view of a phosphorene sheet encapsulated by media with dielectric tensors $\boldsymbol{\varepsilon}_a$ and $\boldsymbol{\varepsilon}_b$. (b) top view indicating the electric field orientation.  }\label{fig:geometry}
\end{figure}
By now, it is well known that many excitonic effects are accurately described by modeling the excitons as electron-hole pairs satisfying the Wannier equation \cite{Wannier1937,Lederman1976}. It has been shown repeatedly that the Wannier model reproduces the exciton binding energies obtained from first principle calculations to a satisfactory degree \cite{Cudazzo2010,Pulci2012,Latini2015,Henriques2019}. This is fortunate, as first principle calculations require solving the computationally demanding Bethe-Salpeter equation \cite{Salpeter1951,Onida2002}. Importantly, the Wannier model has also been shown to agree with experimental results for the exciton Stark effect, field-induced dissociation, and the Franz-Keldysh effect \cite{Massicotte2018}. For anisotropic 2D semiconductors, the Wannier equation in the absence of an electric field reads (in atomic units)
\begin{align}
	\left[-\frac{1}{2\mu_x}\frac{\partial^2}{\partial x^2} 	-\frac{1}{2\mu_y}\frac{\partial^2}{\partial y^2} + V\left(\boldsymbol{r}\right) - E\right]\psi\left(\boldsymbol{r}\right)=0\thinspace,\label{eq:Wannier}
\end{align}
where $V$ is the electron-hole interaction, $E$ the energy, and 
\begin{align}
	\mu_{x/y} = \frac{m_e^{\left(x/y\right)}m_h^{\left(x/y\right)}}{m_e^{\left(x/y\right)}+m_e^{\left(x/y\right)}}
\end{align}
is the direction-specific reduced mass with $m_e^{\left(x/y\right)}$ and $m_h^{\left(x/y\right)}$ being the electron and hole effective masses along the $x/y$-direction, respectively. The exciton interaction $V$ in anisotropic semiconductors is, of course, anisotropic. It may be found by modeling the encapsulated 2D semiconductor  (depicted in \cref{fig:geometry} (a) ) as a three-layer structure with a piecewise constant dielectric function, and then solving the Poisson equation for two charges in this structure. This is done in \cref{app:poisson}. For a superstrate, a 2D semiconductor, and a substrate with dielectric tensors $\varepsilon_a = \mathrm{diag}(\varepsilon_{xx}^{\left(a\right)},\varepsilon_{xx}^{\left(a\right)},\varepsilon_{zz}^{\left(a\right)})$, $\varepsilon = \mathrm{diag}(\varepsilon_{xx},\varepsilon_{yy},\varepsilon_{zz})$, and $\varepsilon_b = \mathrm{diag}(\varepsilon_{xx}^{\left(b\right)},\varepsilon_{xx}^{\left(b\right)},\varepsilon_{zz}^{\left(b\right)})$, respectively, we find with a linearized dielectric function
\begin{multline}
V\left(r,\theta\right) \approx-\int_{0}^\infty\frac{J_0\left(qr\right)}{\sqrt{\varepsilon_x\varepsilon_y}}dq \\-2\sum_{k=1}^{\infty}\cos\left(2k\theta\right)\int_{0}^\infty\left(\frac{\sqrt{\varepsilon_x}-\sqrt{\varepsilon_y}}{\sqrt{\varepsilon_x}+\sqrt{\varepsilon_y}}\right)^kJ_{2k}\left(qr\right)dq\thinspace,\label{eq:fullpotential}
\end{multline}
where 
$\varepsilon_x = \kappa + r_{0x}q$, $\varepsilon_y = \kappa + r_{0y}q$, and $\kappa = \frac{\sqrt{\varepsilon_{xx}^{\left(a\right)}\varepsilon_{zz}^{\left(a\right)}}+\sqrt{\varepsilon_{xx}^{\left(b\right)}\varepsilon_{zz}^{\left(b\right)}}}{2}$. The screening lengths are defined as $r_{0x}=2\pi\alpha_{xx}^{\left(\mathrm{2D}\right)}$ and $r_{0y}=2\pi\alpha_{yy}^{\left(\mathrm{2D}\right)}$, where $\alpha_{xx}^{\left(\mathrm{2D}\right)}$ and $\alpha_{yy}^{\left(\mathrm{2D}\right)}$ are the 2D sheet polarizabilities in the $x$- and $y$-direction, respectively. These are the microscopic definitions of the screening lengths \cite{Berkelbach2013}. The macroscopic definitions may be seen in \cref{app:poisson}. Note that \cref{eq:fullpotential} reduces to the usual Rytova-Keldsyh potential \cite{Rytova1967,Keldysh1979,Trolle2017} in the isotropic case. 

The polarizabilities for phosphorene were computed in Ref. \cite{Rodin2014}, where the authors found $\alpha_{xx}^{\left(\mathrm{2D}\right)} = \SI{4.20}{\angstrom}$ and $\alpha_{yy}^{\left(\mathrm{2D}\right)} = \SI{3.97}{\angstrom}$. These values are quite close, which in turn leads to a very weak angular contribution to the interaction. As a first approximation, we may therefore consider only the leading term. Further expanding $\sqrt{\varepsilon_x\varepsilon_y}$ to first order in $q$, we find the Rytova-Keldysh form \cite{Rytova1967,Keldysh1979}
\begin{align}
	V\left(r,\theta\right)\approx V_{\mathrm{RK}}\left(r\right) = -\frac{\pi}{2r_0}\left[\mathrm{H}_0\left(\frac{\kappa r}{r_0}\right)-Y_0\left(\frac{\kappa r}{r_0}\right)\right]\thinspace,\label{eq:RK}
\end{align}
\begin{figure}[t]
	\includegraphics[width=1\columnwidth]{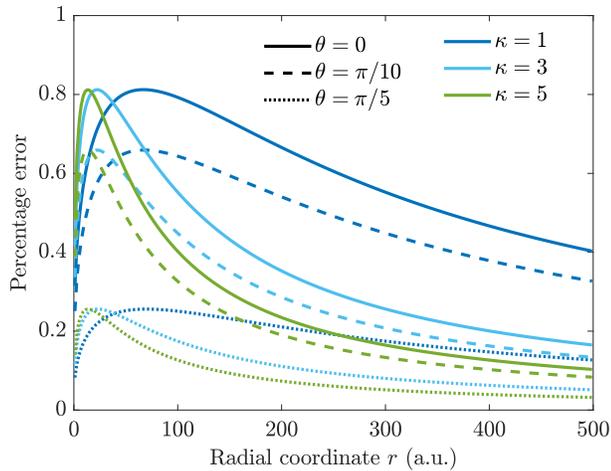}
	\caption{Error introduced by using the isotropic approximation in \cref{eq:RK} instead of the full interaction in \cref{eq:fullpotential} to describe excitons in phosphorene.  }\label{fig:potential_error}
\end{figure}
with $r_0 = \left(r_{0x}+r_{0y}\right)/2$. This form agrees with the interaction used by many authors to study excitons in phosphorene \cite{Rodin2014,Junior2019,Henriques2020}. Using the isotropic approximation to describe the excitons is justified by noting that it agrees with the full potential in \cref{eq:fullpotential} to within $1\%$ for all $r$ and $\theta$. Errors for three different dielectric surroundings and field directions are shown in \cref{fig:potential_error}. As is evident, these errors are low, and they decrease as $\theta$ tends to $\pi/4$, as the dominating angular contribution from the $k=1$ term tends to zero in this region. The parameters used to describe excitons in phosphorene in the present paper are summarized in \cref{tbl:parameters}.
\begingroup
\squeezetable
\begin{table}[b]
	\caption{Parameters used to describe excitons in phosphorene. The effective masses are from Ref. \cite{Choi2015}, the polarizabilities are from Ref. \cite{Rodin2014}, and the rest of the parameters are computed from these values.  }
	\centering
	\begin{tabular}{cc cc cc cc cc cc cc cc c}
		\hline\hline\\[-0.2cm]
		$\alpha_{xx}^{\left(\mathrm{2D}\right)}$ & \phantom & $\alpha_{yy}^{\left(\mathrm{2D}\right)}$ & \phantom & $m_e^{\left(x\right)}$  &\phantom
		& $m_e^{\left(y\right)}$  & \phantom &$m_h^{\left(x\right)}$ & \phantom &$m_h^{\left(y\right)}$ &\phantom & $r_0$ & \phantom& $\mu_x$ & \phantom &$\mu_y$ \\[0.1cm]\hline \\[-0.2cm]
		$\SI{4.20}{\angstrom}$&\phantom & $\SI{3.97}{\angstrom}$& \phantom &$0.46$ &\phantom& $1.12$ & \phantom & $0.23$ & \phantom &$1.61$ & \phantom & $\SI{25.67}{\angstrom}$ &\phantom& $0.1533$ & \phantom &  $0.6605$\\[0.1cm]
		%Entering 1st row
	\end{tabular}\label{tbl:parameters}
\end{table} 
\endgroup
\begin{table}[b]
	\caption{Symmetry of the four types of unperturbed states. The first column indicates the axis or point considered and the remaining columns indicate the symmetries or properties of the states about said axis or point. The final two rows indicate whether or not the states have angular nodes along the $X$- or $Y$-axis. A zero indicates an angular node, while a dash indicates that nothing general may be inferred about the states along the axes from their symmetry. }
	\centering
	\begin{tabular}{c c c c c c c c c c c}
		\hline\hline\\[-0.2cm]
		\phantom &\multicolumn{1}{ c }{State $ce$} 
		&\phantom&\multicolumn{1}{ c }{State $co$}
		&\phantom&\multicolumn{1}{ c }{State $se$}&\phantom&\multicolumn{1}{ c }{State $so$}\\ [0.1cm]\hline\\[-0.2cm]
		About $X$-axis & Even &\phantom& Even &\phantom& Odd &\phantom & Odd \\[0.1cm]
		About $Y$-axis & Even &\phantom& Odd &\phantom& Odd & \phantom & Even\\[0.1cm]
		About origin & Even &\phantom& Odd &\phantom& Even & \phantom & Odd\\[0.1cm]
		Along $X$-axis & - &\phantom& - &\phantom& 0 & \phantom & 0\\[0.1cm]
		Along $Y$-axis & - &\phantom& 0 &\phantom& 0 & \phantom & -
		%Entering 1st row
		
	\end{tabular}\label{tbl:symmetry}
\end{table} 

To perform numerical calculations, it is convenient to switch to the coordinates introduced in Ref. \cite{Rodin2014} defined by 
\begin{align}
	X = \sqrt{\frac{\mu_x}{2\mu}}x\thinspace, \quad Y = \sqrt{\frac{\mu_y}{2\mu}}y\thinspace, \quad \mu = \frac{\mu_x\mu_y}{\mu_x+\mu_y}\thinspace.
\end{align}
This transforms \cref{eq:Wannier} into
\begin{align}
	\left\{-\frac{1}{4\mu}\nabla^2 + V_{\mathrm{RK}}\left[R\sqrt{1+\beta\cos\left(2\Theta\right)}\right]-E\right\}\psi\left(\boldsymbol{R}\right) = 0\thinspace,\label{eq:Wanniermod}
\end{align}
with $\beta = \left(\mu_y-\mu_x\right)/\left(\mu_y+\mu_x\right)$, $R$ and $\Theta$ the polar representation of the $XY$-plane, and where the Laplacian is to be taken with respect to these coordinates. The transformation makes the kinetic energy isotropic at the cost of making the potential energy anisotropic. The reasons that this transformation is useful are threefold: firstly, it is more intuitive to work with an anisotropic potential than an anisotropic kinetic energy; secondly, the numerical procedure we shall use consists of resolving the wave function in a basis, and the number of basis functions needed to represent the polar wave function is significantly less than those needed to represent the Cartesian wave function; and, finally, the polar representation of the kinetic energy is simple, and thus leads to simple matrix elements. The states may be expressed generally as
\begin{align}
	\psi\left(\boldsymbol{R}\right) =  \sum_{m=-\infty}^{\infty}\varphi_m\left(R\right)\times \begin{cases}
	\cos\left(m\Theta\right)\thinspace, \quad m\leq0\\
	\sin\left(m\Theta\right)\thinspace, \quad m>0\thinspace,
	\end{cases}\label{eq:expansion}
\end{align}
where $\varphi_m$ may be understood as the Fourier coefficients. Note that the coefficients depend continuously on $R$, and will later be expanded in a radial basis. As pointed out in Ref. \cite{Rodin2014}, the unperturbed eigenstates of \cref{eq:Wanniermod} fall into four distinct groups. This is most easily seen by recognizing that the potential is even in $\Theta$ and is invariant under the shift $\Theta \to \Theta + \pi$. It therefore has a Fourier series consisting of cosines of even order, and it becomes easy to see that for coupling to occur, the angular functions must be of the same type (i.e. sine/cosine) and have the same angular momentum parity. This results in the following four types of states
\begin{align}
	&\psi_{ce} = \sum_{m=0}^\infty \varphi_m^{\left(ce\right)}\left(R\right) \cos\left(2m\Theta\right)\thinspace,\label{eq:state1}\\
	&\psi_{co} = \sum_{m=0}^\infty \varphi_m^{\left(co\right)}\left(R\right) \cos\left[\left(2m+1\right)\Theta\right]\thinspace,\label{eq:state2}\\
	&\psi_{se} = \sum_{m=1}^\infty \varphi_m^{\left(se\right)}\left(R\right) \sin\left(2m\Theta\right)\thinspace,\label{eq:state3}\\
	&\psi_{so} = \sum_{m=0}^\infty \varphi_m^{\left(so\right)}\left(R\right) \sin\left[\left(2m+1\right)\Theta\right]\thinspace,\label{eq:state4}
\end{align}
where the subscripts $c/s$ and $e/o$ denote whether the trigonometric function is a cosine/sine and whether the angular momentum parity is even/odd, respectively. A couple of symmetry observations follow immediately and are summarized in \cref{tbl:symmetry}. Importantly, states with even and odd angular momenta are symmetric and antisymmetric about the origin, respectively. This means that the $co$ and $so$ states are necessarily zero at the origin, and are therefore not optically active when no external electric field is present, as we shall see later on when we compute the optical absorption. The shape of the states may be inferred from the fact that some of them have angular nodes along a specific axis. In particular, the $co$ and $so$ states have nodes along the $Y$- and $X$-axis, respectively, and the $se$ state has nodes along both. They will therefore be slightly deformed versions of the familiar $p_x$, $p_y$, and $d_{xy}$ orbitals. 

The numerical procedure we will use throughout the paper is to solve the Wannier equation using a finite element representation of the exciton wave functions. The wave functions are expanded as in \cref{eq:expansion}, including angular momenta up to some number $M$. The radial $\varphi_m$ functions are then expanded in a finite element basis $f_i^{\left(n\right)}$ that are non-zero only on a single radial segment $n$ defining a particular range of the radial coordinate. Their exact form, as well as further detail of the numerical procedure can be found in Ref. \cite{Kamban2019}. To summarize,
\begin{align}
	\varphi_m\left(\Theta\right) = \sum_{n=1}^{N}\sum_{i=1}^{p}c_i^{\left(m,n\right)}f_i^{\left(n\right)}\left(R\right)\thinspace,
\end{align}
where $c_i^{\left(m,n\right)}$ are the expansion coefficients obtained by solving the resulting matrix eigenvalue problem.
\begin{table}[b]
	\caption{Five lowest exciton energies in phosphorene in three different dielectric surroundings. The first column numerates the states with increasing energy, the second column indicates which of the four types (\cref{eq:state1,eq:state2,eq:state3,eq:state4}) the state belongs to, and the remaining columns show the exciton energies in the three dielectric surroundings. }
	\centering
	\begin{tabular}{c c c c c c c c}
		\hline\hline\\[-0.2cm]
		\phantom & \phantom &\multicolumn{1}{ c }{Freely suspended} 
		&\phantom&\multicolumn{1}{ c }{SiO$_2$ substrate}
		&\phantom&\multicolumn{1}{ c }{hBN encapsulation}\\
		\cmidrule{3-3}  \cmidrule{5-5} \cmidrule{7-7}\\[-0.2cm]
		$n$ & Type &$E_n$ (meV)  & \phantom 
		&$E_n$ (meV)   &\phantom
		&$E_n$ (meV)   		\\ [0.1cm]\hline\\[-0.2cm]
		$0$ &$ce$ & $-822$ &\phantom& $-459 $ &\phantom& $-260 $ \\[0.1cm]
		$1$ &$so$ & $-519 $ &\phantom& $-227$ &\phantom& $-99 $\\[0.1cm]
		$2$ &$ce$ & $-410 $ &\phantom& $-163$ &\phantom& $-67 $\\[0.1cm]
		$3$ &$co$ & $-385$ &\phantom& $-145$ &\phantom& $-55$\\[0.1cm]
		$4$ &$so$ & $-320$ &\phantom& $-113$ &\phantom& $-42$ \\[0.1cm]
		%Entering 1st row
		
	\end{tabular}\label{tbl:energy}
\end{table} 

In the present paper, we will study phosphorene in three different dielectric environments. Namely, freely suspended ($\kappa=1$), on an SiO$_2$ substrate ($\kappa = 2.4$), and encapsulated by hBN ($\kappa = 4.5$). The energy and symmetry of the five lowest lying states for the three dielectric surroundings are shown in \cref{tbl:energy}. As can be seen, we find exciton binding energies of $822$, $459$, and $260$ meV, respectively, which corresponds well with those presented in Refs. \cite{Rodin2014,Chaves2015,Yang2015,Hunt2018,Junior2019,Henriques2020} (see Ref. \cite{Henriques2020} for a table summarizing the binding energies from more references). Additionally, the energies of the excited states are in good agreement with those found in Ref. \cite{Chaves2015}. Note that the $n=1$ state is an $so$ state that has an angular node along the $X$-axis. The fact that this state has a lower energy than the $co$ state can be understood by considering the potential in \cref{eq:Wanniermod}. It is weaker along the $X$-axis than it is along the $Y$-axis, thus favoring a state along $Y$. As a final note, the $se$ state is not among the first five states. That it has such a large energy is no surprise given its $d_{xy}$-like shape.

\section{Anisotropic exciton Stark effect}\label{sec:Stark}

We are interested in seeing how the anisotropic nature of phosphorene affects both the exciton Stark shifts and dissociation rates when an in-plane electrostatic field is applied to the sheet. In the presence of an electric field, the Wannier equation reads
\begin{multline}
	\left\{-\frac{1}{4\mu}\nabla^2 + V\left[R\sqrt{1+\beta\cos\left(2\Theta\right)}\right]+\mathcal{E}\cos\alpha\,\sqrt{\frac{2\mu}{\mu_x}}X\right.\\\left.+\,\mathcal{E}\sin\alpha\,\sqrt{\frac{2\mu}{\mu_y}}Y-E\right\}\psi\left(\boldsymbol{R}\right) = 0\thinspace,\label{eq:Wannierwfield}
\end{multline} 
where $\mathcal{E}$ is the electric field strength and $\alpha\in\left[0,\pi/2\right]$ its angle to the original $x$-axis. The setup is pictured in \cref{fig:geometry}, where the field direction is indicated in panel (b). Applying a field to the exciton induces resonance states and, in turn, makes the energy eigenvalue complex \cite{Haastrup2016,Pedersen2016ExcitonIonization}. The Stark shift then corresponds to the change in the real part of the energy as a function of field strength, while the imaginary part describes the dissociation rate by the relation $\Gamma = -2\,\mathrm{Im}E$. This complex eigenvalue is most easily obtained by using the complex scaling procedure \cite{Balslev1971,Aguilar1971}. Here, the radial coordinate is rotated into the complex plane, which transforms the diverging behavior of the resonance states for real $r$ into bound states along a complex contour $re^{i\varphi}$. 
\begin{figure}[t]
	\includegraphics[width=1\columnwidth]{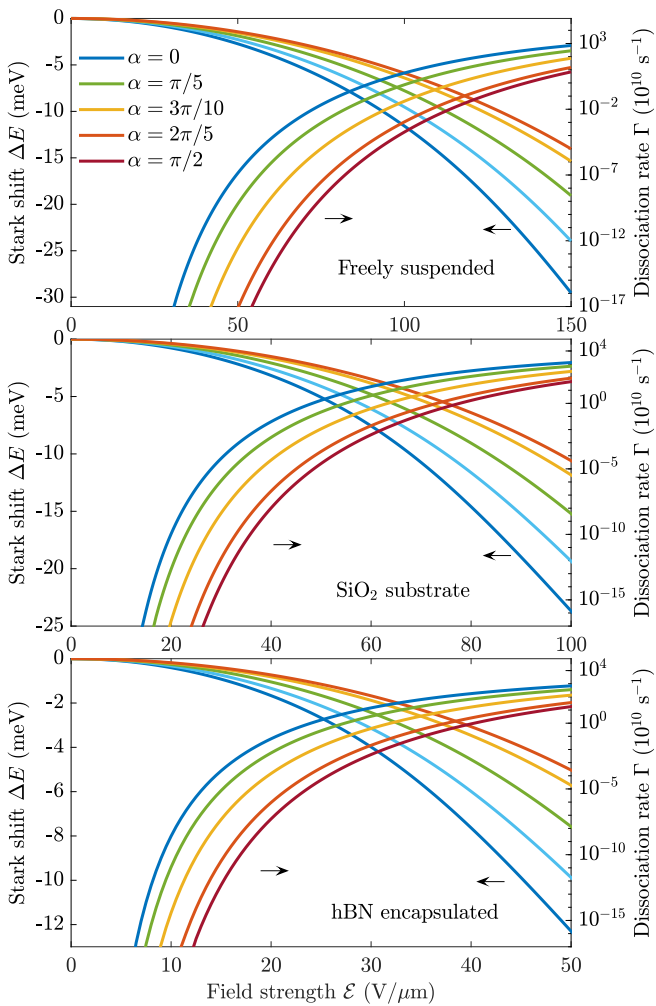}
	\caption{Stark shifts $\Delta E = \mathrm{Re}\left(E-E_0\right)$ (left axis) and dissociation rates $\Gamma = -2\mathrm{Im}\,E$ (right axis) of excitons in phosphorene that is freely suspended (top), on an SiO$_2$ substrate (middle), or encapsulated by hBN (bottom). For each of the three cases, electric fields with an angle to the $x$-axis $\alpha$ ranging from $0$ to $\pi/2$ are considered. Evidently, the Stark shifts and dissociation rates decrease with an increasing angle, as expected, since the $y$-component of the reduced mass is much larger than the $x$-component. }\label{fig:SS_and_diss}
\end{figure}

The procedure we shall use here has been laid out in Ref. \cite{Kamban2019}, where we compute the exciton Stark effect in TMDs. It consists of expanding the resonance state in a finite element basis, as described above, and only complex scaling the coordinate outside a desired radius. This technique is referred to as exterior complex scaling \cite{Scrinzi1993}, and it makes it much easier to obtain the dissociation rates for weak fields numerically. The results for phosphorene in free space, on an SiO$_2$ substrate, and encapsulated by hBN can be seen in the top, middle, and bottom panel of \cref{fig:SS_and_diss}, respectively. It is immediately clear that the field direction, indicated by the line color, is very important. This is in contrast to the effect in TMDs, which is fully isotropic \cite{Haastrup2016,Massicotte2018,Kamban2019}. The largest Stark shifts (left axis) and dissociation rates (right axis) are seen for fields pointing along the $x$-axis, which coincides with the direction of lowest effective mass. Rotating the field from the $x$-axis to the $y$-axis reduces the dissociation rate by several orders of magnitude, due to the increased effective mass. The direction dependent Stark shifts and dissociation rates add an additional degree of freedom when using phosphorene in device design, as not only can they be controlled by the dielectric environment, but by the field direction as well. Taking a closer look at the effect of the dielectric surroundings, both the Stark shifts and dissociation rates increase significantly by placing the phosphorene sheet on an SiO$_2$ substrate, and even more so by encapsulating it in hBN. This is to be expected, as the exciton binding energy is reduced considerably with increased screening. It should also be noted that the shifts and rates are much lower than those in popular TMDs \cite{Kamban2019}, which is a direct consequence of the larger binding energies of excitons in phosphorene.

\subsection{Exciton Stark shift and polarizability}

\begin{figure}[t]
	\includegraphics[width=1\columnwidth]{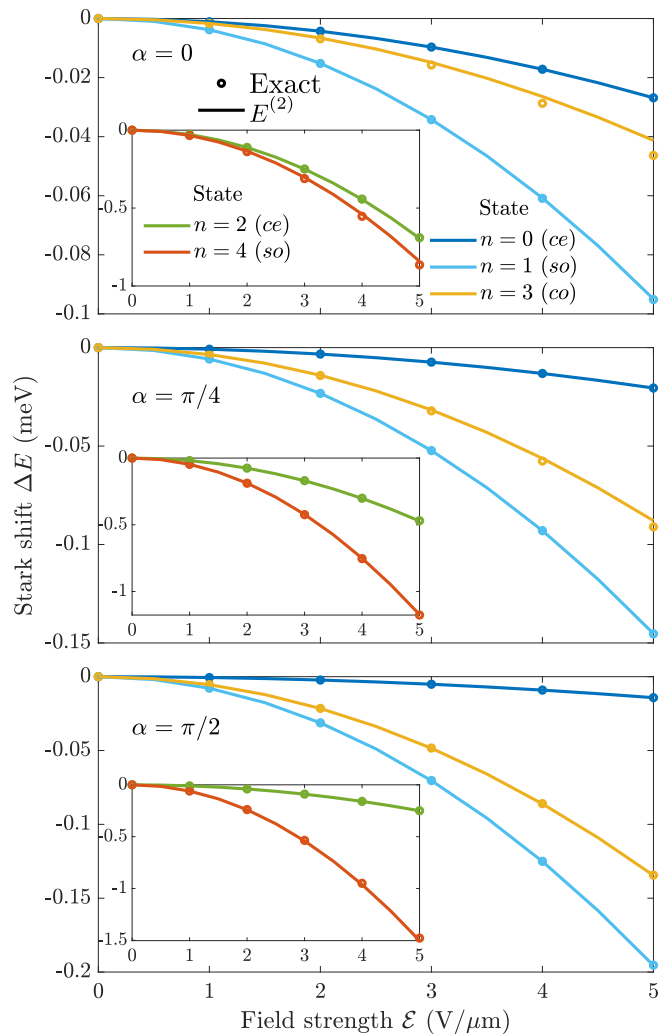}
	\caption{Comparison between the numerically exact Stark shifts (circles) and second order perturbation theory \cref{eq:E2} (solid lines) for the five lowest lying exciton states in freely suspended phosphorene. States $n\in\left\{0,1,3\right\}$ are shown in the large panels, while the insets show $n\in\left\{2,4\right\}$. The top, middle, and bottom panels represent fields with an angle to the $x$-axis of $0$, $\pi/4$, and $\pi/2$, respectively. Good agreement is found across all field angles for very weak electric fields. }\label{fig:SS}
\end{figure}

In this section, we shall look at the exciton Stark shift in more detail, and compare to the shift predicted by perturbation theory. The anisotropic exciton Stark shift in few layer BP in free space and in hBN surroundings was studied in Ref. \cite{Cavalcante2018}, and on an SiO$_2$ substrate in Ref. \cite{Chaves2015}, and we shall thus make a detailed comparison to the results in these papers for a single layer of BP, i.e. phosphorene. One of the most important quantities describing how anisotropic excitons interact with the electric field is their polarizabilities. The perturbation series for the Stark shift of state $n$ may be written as \cite{Landau1989}
\begin{align}
	\Delta E_n = \mathrm{Re}\left[E_n\left(\mathcal{E}\right)-E_n^{\left(0\right)}\right] = E_n^{\left(1\right)} +  E_n^{\left(2\right)} + O\left(\mathcal{E}^3\right)\thinspace,
\end{align}
with 
\begin{align}
	E_n^{\left(1\right)} = \bra{\psi_n^{\left(0\right)}}H'\ket{\psi_n^{\left(0\right)}}
\end{align}
and
\begin{align}
	E_n^{\left(2\right)} = \sum_{k\neq n}\frac{\left|\bra{\psi_k^{\left(0\right)}}H'\ket{\psi_n^{\left(0\right)}}\right|^2}{E_n^{\left(0\right)}-E_k^{\left(0\right)}}
\end{align}
where 
\begin{align}
	H' = \mathcal{E}\cos\alpha\,\sqrt{\frac{2\mu}{\mu_x}}X+\,\mathcal{E}\sin\alpha\,\sqrt{\frac{2\mu}{\mu_y}}Y\thinspace,
\end{align}
and the sum is to be taken over all the unperturbed states. The perturbation $H'$ only couples states with different parity angular momenta (see \cref{eq:state1,eq:state2,eq:state3,eq:state4}). The first order correction therefore immediately reduces to zero. The second order correction, on the other hand, may be written as
\begin{align}
E_n^{\left(2\right)} =  -\frac{1}{2}\chi_{n}^{XX}\cos^2\alpha\,\mathcal{E}^2-\frac{1}{2}\chi_{n}^{YY}\sin^2\alpha\,\mathcal{E}^2\thinspace,\label{eq:E2}
\end{align}
where the $X$- and $Y$-components of the exciton polarizability tensor for the four types of states are given by
\begin{align}
	\chi_{ce}^{XX} = \frac{4\mu}{\mu_x}\sum_{co}\frac{\left|X_{co,ce}\right|^2}{E_{co,ce}}\thinspace, \,\,\,\chi_{ce}^{YY} = \frac{4\mu}{\mu_y}\sum_{so}\frac{\left|Y_{so,ce}\right|^2}{E_{so,ce}}\\
	\chi_{co}^{XX} = \frac{4\mu}{\mu_x}\sum_{ce}\frac{\left|X_{ce,co}\right|^2}{E_{ce,co}}\thinspace, \,\,\,\chi_{co}^{YY} = \frac{4\mu}{\mu_y}\sum_{se}\frac{\left|Y_{se,co}\right|^2}{E_{se,co}}\\
	\chi_{se}^{XX} = \frac{4\mu}{\mu_x}\sum_{so}\frac{\left|X_{so,se}\right|^2}{E_{so,se}}\thinspace, \,\,\,\chi_{se}^{YY} = \frac{4\mu}{\mu_y}\sum_{co}\frac{\left|Y_{co,se}\right|^2}{E_{co,se}}\\
	\chi_{so}^{XX} = \frac{4\mu}{\mu_x}\sum_{se}\frac{\left|X_{so,se}\right|^2}{E_{so,se}}\thinspace, \,\,\,\chi_{so}^{YY} = \frac{4\mu}{\mu_y}\sum_{ce}\frac{\left|Y_{ce,so}\right|^2}{E_{ce,so}}
\end{align}
with the shorthand notation
\begin{align}
	X_{ij}=\bra{\psi_{i}^{\left(0\right)}}X\ket{\psi_{j}^{\left(0\right)}}\thinspace&,\quad Y_{ij}=\bra{\psi_{i}^{\left(0\right)}}Y\ket{\psi_{j}^{\left(0\right)}}\thinspace,\\
	&E_{ij}= E_i^{\left(0\right)}-E_j^{\left(0\right)}\thinspace.
\end{align}
The finite element expansion described above is very flexible and perfectly capable of resolving both the bound and continuous spectrum of unperturbed states. In this manner, we compute the exciton polarizabilities for the five lowest lying exciton states. They are summarized in \cref{tbl:polarizability}, and a comparison between \cref{eq:E2} and the numerically exact Stark shifts is shown in \cref{fig:SS} for very weak fields. It is clear that all of the states have a highly anisotropic response to an applied field. We observe, as expected, that the $X$-component of the polarizability is larger than the $Y$-component for the fundamental exciton. The opposite is the case for the $so$ states, which is of no surprise given their $p_y$-like shape. Perhaps more surprisingly we find that the $Y$-component of the $co$ state is larger than the $X$-component. 
\begin{table}[b]
	\caption{Exciton polarizabilities in units of $10^{-18}\,\mathrm{eV}(\mathrm{m}/\mathrm{V})^2$ of the five lowest lying excitons in phosphorene in three different dielectric surroundings. The first column numerates the states with increasing energy (shown in \cref{tbl:energy}), the second column indicates which of the four types (\cref{eq:state1,eq:state2,eq:state3,eq:state4}) the states belong to, and the remaining columns show the exciton polarizabilities in the three dielectric surroundings. }
	\centering
	\begin{tabular}{cc cc c cc c cc}
		\hline\hline\\[-0.2cm]
		\phantom & \phantom &\multicolumn{2}{ c }{Freely suspended} 
		&\phantom&\multicolumn{2}{ c }{SiO$_2$ substrate}
		&\phantom&\multicolumn{2}{ c }{hBN encapsulation}\\
		\cmidrule{3-4}  \cmidrule{6-7} \cmidrule{9-10}\\[-0.2cm]
		$n$ & Type &$\chi^{XX}_n$ & $\chi^{YY}_n$& \phantom 
		&$\chi^{XX}_n$ & $\chi^{YY}_n$   &\phantom
		&$\chi^{XX}_n$ & $\chi^{YY}_n$\\[0.1cm]
		\hline\\[-0.2cm]
		0 & $ce$ & $2.149$ & $1.137$ & \phantom & $3.830$ & $1.870$ & \phantom & $8.058$ & $3.614$ \\[0.1cm]
		1 & $so$ & $7.603$ & $15.65$ & \phantom & $20.01$ & $43.69$ & \phantom & $71.08$ & $168.7$ \\[0.1cm]
		2 & $ce$ & $55.57$ & $20.02$ & \phantom & $184.9$ & $59.49$ & \phantom & $749.1$ & $195.8$ \\[0.1cm]
		3 & $co$  & $3.303$ & $10.77$ & \phantom & $18.46$ & $34.70$ & \phantom & $184.9$ & $158.5$ \\[0.1cm]
		4 & $so$ & $67.42$ & $120.3$ & \phantom & $290.0$ & $583.5$ & \phantom & $1559$ & $3673$ \\[0.1cm]   
		%Entering 1st row
		
	\end{tabular}\label{tbl:polarizability}
\end{table} 

In Ref. \cite{Chaves2015}, the authors use the RK potential with parameters that are almost identical to ours to study excitons in phosphorene on an SiO$_2$ substrate. However, our reduced masses are slightly different from theirs. Specifically, the authors use $\mu_x = 0.089$ and $\mu_y = 0.650$ from Ref. \cite{Castellanos-Gomez2014} instead of our $\mu_x = 0.153$ and $\mu_y = 0.661$ from Ref. \cite{Choi2015}. We therefore expect our exciton energies to be slightly lower, as well as a lower degree of anisotropy. Indeed, they find $E_0 = -396$ and $E_2 = -143$ meV, where we find $E_0 = -459$ and $E_2 = -163$ meV, showing a good qualitative agreement. Turning to the polarizabilities, the authors of Ref. \cite{Chaves2015} determine the polarizabilities of the $n=0$ and $n=2$ states by fitting to the numerical Stark shifts. Using this procedure, the authors find the polarizabilities in the $X\,(Y)$-direction to be $7.4\,\left(2.7\right)$ and $200\,\left(96\right)$ for the $n=0$ and $n=2$ exciton, respectively, all in units of $10^{-18}\,\mathrm{eV}(\mathrm{m}/\mathrm{V})^2$. Comparing to our results in  \cref{tbl:polarizability}, we find $3.8\,\left(1.9\right)$ and $185\,\left(59.5\right)$ for the same cases. We thus obtain slightly lower polarizabilities as expected from the larger binding energies. The same authors also compute the Stark shifts of the fundamental exciton for field strengths up to $20\,$ V/$\mu$m, and find shifts of around $1.6$ and $0.6$ meV for fields in the $X$- and $Y$-direction, respectively. For the same fields, we find shifts of $0.8$ and $0.4$ meV. In Ref. \cite{Cavalcante2018}, the authors study the Stark shifts for freely suspended phosphorene, as well as phosphorene in hBN surroundings. These authors use the same reduced mass as the authors of Ref. \cite{Chaves2015}, and find polarizabilities of $1.861 \,(0.871)$ and $6.162\, (2.598)\times 10^{-18}\mathrm{eV}\left( \mathrm{m}/\mathrm{V}\right)^2$ in the $X\,(Y)$-direction for freely suspended and hBN encapsulated phosphorene, respectively. These values are in good agreement with our results. 

\subsection{Analytical weak-field approximation for exciton dissociation}

Analytical weak-field expressions for the dissociation rates of excitons in monolayer \cite{Kamban2019,Henriques2020b} and bilayer TMDs \cite{Kamban2020} have been derived previously using weak-field asymptotic theory (WFAT) \cite{Tolstikhin2011}. These expressions are useful for obtaining a better understanding of field-induced exciton dissociation, as well as obtaining quick estimates of the rates at weak electric fields without performing heavy numerical computations. This latter point is of great importance, as the numerical procedures break down for sufficiently weak fields \cite{Trinh2013}. To implement WFAT, the binding potential must have a sufficiently simple asymptotic behavior (specified later), and the electric field must point along the $x$-axis (or $z$-axis for 3D problems). The reason that the field must point along this axis is that it is very simple to deal with in parabolic coordinates, and leads to differential equations that decouple in the asymptotic region. For TMDs, advantage was taken of the isotropic nature of the problem by letting the electric field point in the $x$-direction, and the resulting problem therefore had the form
\begin{align}
	\left[-\frac{1}{2}\nabla^2 + V\left(r\right) + \mathcal{E}x-E\right]\psi\left(\boldsymbol{r}\right)=0\thinspace,\label{eq:stdform}\\
	V\left(r\right) \sim -\frac{1}{\kappa r} \quad \text{for } r\to\infty\notag
\end{align}
where $V$ is a radial potential. Parabolic coordinates thus allow separation of this problem in the asymptotic region. A similar procedure may be used in the present case, but a coordinate transformation is needed in order to get the desired axis to coincide with the field direction.

Generalizing the coordinate transformations in the previous section slightly, we write
\begin{align}
\zeta_x &= \sqrt{\mu_x}\cos\tilde{\alpha}\,x+\sqrt{\mu_y}\sin\tilde{\alpha}\,y\\
\zeta_y &= -\sqrt{\mu_x}\sin\tilde{\alpha}\,x+\sqrt{\mu_y}\cos\tilde{\alpha}\,y\thinspace.
\end{align}
Choosing $\tilde{\alpha} = \arctan\left(\sqrt{\frac{\mu_x}{\mu_y}}\tan\alpha\right)$
the Wannier equation for phosphorene becomes
\begin{multline}
\left\{-\frac{1}{2}\nabla^2 + V\left[\frac{\zeta}{\sqrt{2\mu}}\sqrt{1+\beta\cos\left[2\left(\Phi+\tilde{\alpha}\right)\right])}\right] 
\right.\\\left.+ \tilde{\mathcal{E}}\zeta_x - E\right\}\psi\left(\boldsymbol{\zeta}\right)=0\thinspace,\label{eq:scaledWannier}
\end{multline}
with the effective field strength 
\begin{align}
\tilde{\mathcal{E}} = \mathcal{E}\sqrt{\frac{\cos^2\alpha}{\mu_x}+\frac{\sin^2\alpha}{\mu_y}}\thinspace.
\end{align}
Here, $\zeta$ and $\Phi$ are the polar representation of the $\zeta_x$ and $\zeta_y$ coordinates. This brings the equation on the desired form. As shown in \cref{app:WFAT}, applying a weak external field along the $\zeta_x$-axis to a system with an isotropic kinetic energy and an anisotropic potential satisfying
\begin{align}
\lim\limits_{\zeta_x\to-\infty}-\zeta V\left(\boldsymbol{\zeta}\right) = Z_{\mathrm{asymp}}\thinspace,
\end{align}
\begin{figure}[t]
	\includegraphics[width=1\columnwidth]{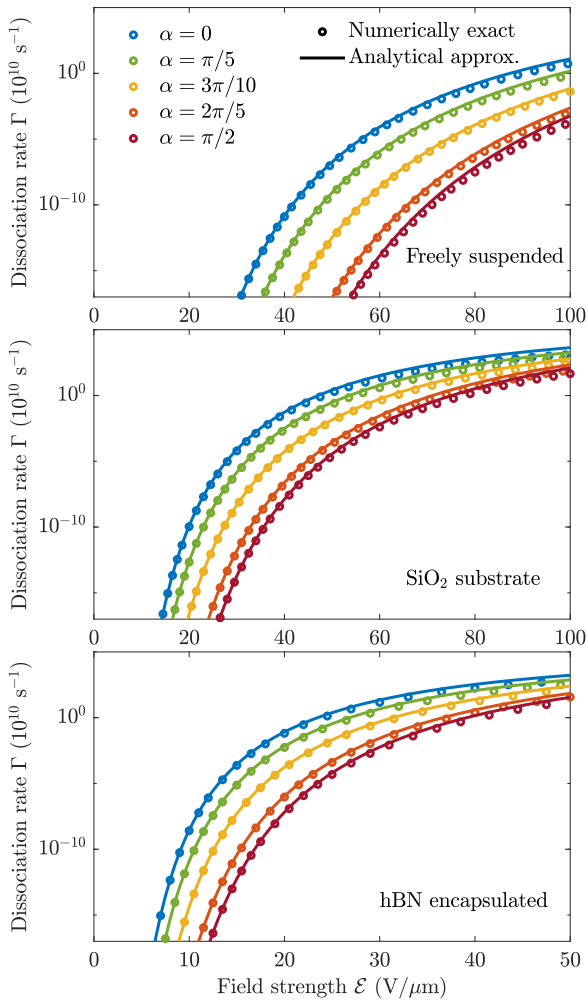}
	\caption{Comparison between the numerically exact dissociation rates (circles) and the anisotropic analytical weak-field approximation in \cref{eq:WFAT} (solid lines). Good agreement is found across all field angles for weak electric fields. }\label{fig:WF_diss}
\end{figure}
\begin{table}[b]
	\caption{Asymptotic coefficient $g_0\left(\alpha\right)$ for phosphorene in three different dielectric surroundings for various field directions. The coefficient is used in the analytical approximation to the exciton dissociation rate \cref{eq:WFAT}, and has been obtained by extrapolating $\Gamma_{\mathrm{exact}}/W_0$ to $\mathcal{E}=0$.}
	\centering
	\begin{tabular}{c c c c c c c}
		\hline\hline\\[-0.2cm]
		\phantom &\multicolumn{1}{ c }{Freely suspended} 
		&\phantom&\multicolumn{1}{ c }{SiO$_2$ substrate}
		&\phantom&\multicolumn{1}{ c }{hBN encapsulation}\\
		\cmidrule{2-2}  \cmidrule{4-4} \cmidrule{6-6}\\[-0.2cm]
		$\alpha$ &$g_0$  & \phantom 
		&$g_0$   &\phantom
		&$g_0$   		\\ [0.1cm]\hline\\[-0.2cm]
		$0$ & $1.835\times 10^{-2}$ & \phantom & $1.344\times 10^{-1}$ & \phantom & $2.122\times 10^{-1}$\\[0.1cm]
		$\frac{\pi}{5}$ & $1.714\times 10^{-2}$ & \phantom & $1.271\times 10^{-1}$ & \phantom & $2.051\times 10^{-1}$\\[0.1cm]
		$\frac{3\pi}{10}$ & $1.239\times 10^{-2}$ & \phantom & $1.090\times 10^{-1}$ & \phantom & $1.812\times 10^{-1}$\\[0.1cm]
		$\frac{2\pi}{5}$ & $2.614\times 10^{-3}$ & \phantom & $4.879\times 10^{-2}$ & \phantom & $9.985\times 10^{-2}$\\[0.1cm]
		$\frac{\pi}{2}$ & $1.495\times 10^{-5}$ & \phantom & $2.892\times 10^{-3}$ & \phantom & $1.322\times 10^{-2}$\\[0.1cm]
		%Entering 1st row
		
	\end{tabular}\label{tbl:1}
\end{table}
\begin{figure*}[t]
	\includegraphics[width=2.1\columnwidth]{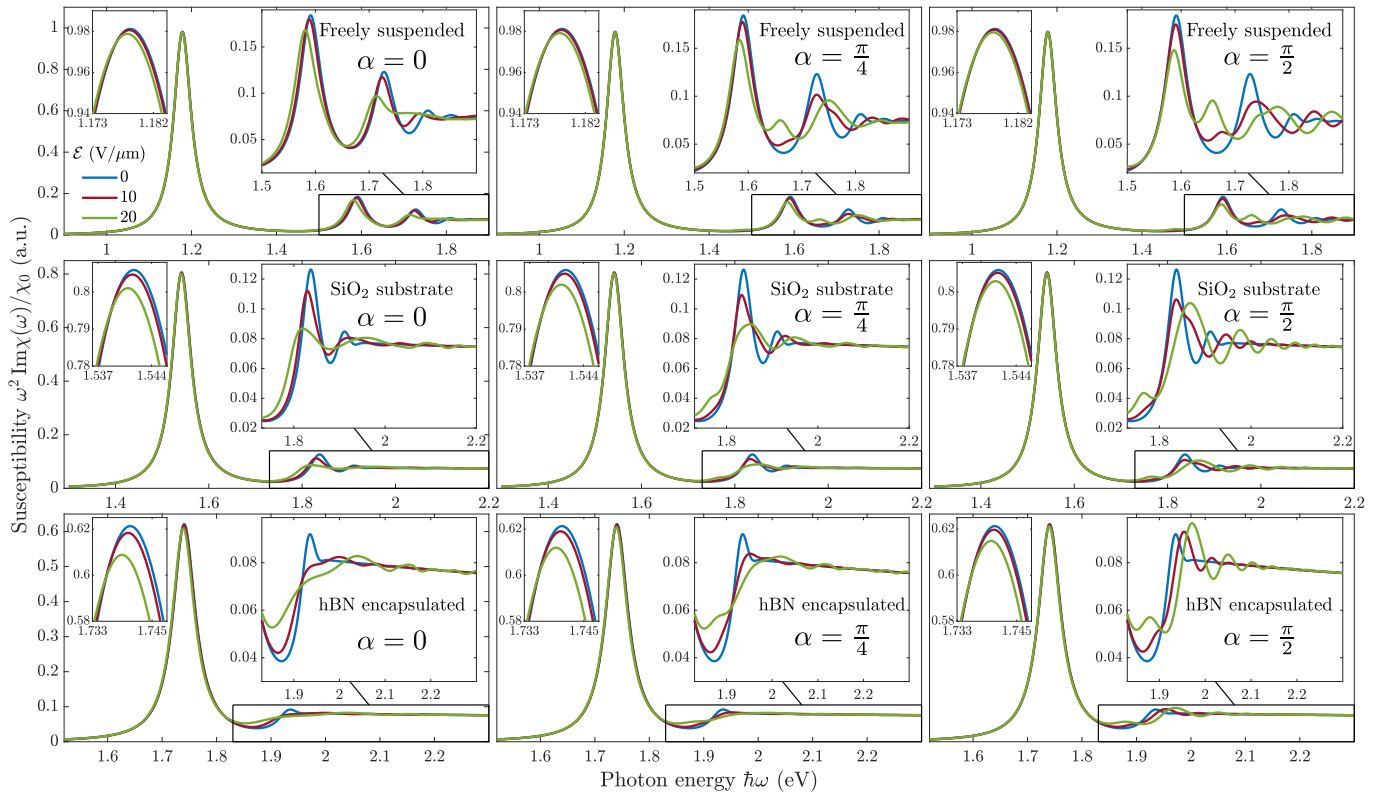}
	\caption{Imaginary part of the susceptibility of phosphorene. The top, middle, and bottom row correspond to phosphorene in free space, on an SiO$_2$ substrate, and encapsulated by hBN, respectively. The first, second, and third column correspond to field angles of $\alpha\in\left\{0,\frac{\pi}{4},\frac{\pi}{2}\right\}$, respectively. Field strengths are indicated by line color, where blue, red, and green represent  $\mathcal{E}\in\left\{0,10,20\right\}$ V/$\mu$m, respectively. Two zooms are shown in each panel: the left is a zoom of the fundamental peak, while the right enhances the spectra at higher photon energies. }\label{fig:Susceptibility_hBN}
\end{figure*}
where $Z_{\mathrm{asymp}}$ is a positive constant, induces the dissociation rate
\begin{align}
\Gamma_{\mathrm{approx}}\left(\mathcal{E},\alpha\right) = \left|g_0\left(\alpha\right)\right|^2W_0\left(\mathcal{E},\alpha\right)\thinspace.\label{eq:WFAT}
\end{align}
Here,
\begin{align}
W_0\left(\mathcal{E},\alpha\right)=k\left(\frac{4k^2}{\tilde{\mathcal{E}}}\right)^{2Z_{\mathrm{asymp}}/k-1/2}\exp\left(- \frac{2k^3}{3\tilde{\mathcal{E}}}\right)
\end{align}
and $k = \sqrt{2|E_0|}$ where $|E_0|$ is the exciton binding energy. For the present case, we find
\begin{align}
Z_{\mathrm{asymp}}\left(\alpha\right) = \frac{1}{\kappa}\sqrt{\frac{2\mu\left(\mu_y\cos^2\alpha+\mu_x\sin^2\alpha\right)}{\left(1+\beta\right)\mu_y\cos^2\alpha+\left(1-\beta\right)\mu_x\sin^2\alpha}}\thinspace.
\end{align}
\Cref{eq:WFAT} generalizes the result in Ref. \cite{Kamban2019} to take into account anisotropic effective masses. In the limit of zero screening length $r_0\to 0$, the RK potential simplifies to the Coulomb potential, and the expression therefore also generalizes the ionization rate of a 2D hydrogen atom \cite{Pedersen2016Stark} to one with an anisotropic reduced mass. Note that it reduces to that of the isotropic case when $\mu_x=\mu_y$. The three direction-dependent quantities in \cref{eq:WFAT} are the field independent asymptotic coefficient $g_0$, the effective field strength $\tilde{\mathcal{E}}$, and $Z_{\mathrm{asymp}}$. The asymptotic coefficient $g_0$ may be computed from the wave function far from the origin (see \cref{eq:g0}). To do so, however, one needs a very accurate numerical wave function. This is reasonably easy to do for the isotropic case, as one may relate the coefficient to the solution of the radial Schr\"odinger equation, obtaining effectively a one-dimensional problem \cite{Kamban2019}. This procedure can not be applied in the present case, and we therefore determine $g_0$ by extrapolating $\Gamma_{\mathrm{exact}}/W_0$ to $\mathcal{E}=0$. The obtained coefficients are shown in \cref{tbl:1}, and the results are compared in \cref{fig:WF_diss}, where the agreement between the numerically exact results and the analytical approximation is good for weak fields. Evidently, the error grows very rapidly with increasing field strength. However, for moderate field strengths, the errors are acceptable, and \cref{eq:WFAT} therefore serves as a decent first approximation to the weak-field exciton dissociation rates in phosphorene.

\section{Franz-Keldysh effect}\label{sec:FK}

The Franz-Keldysh effect \cite{Franz1958,Keldysh1958} constitutes a change in optical absorption of a semiconductor due to an applied external electric field. The effect was computed for monolayer TMDs in Ref. \cite{Pedersen2016ExcitonStark}, and the same methodology used in that paper will be used here. Assuming that the momentum matrix elements are $k$-independent, the exciton oscillator strength may be determined by evaluating the wave function at the origin \cite{Elliott1957}. The exciton susceptibility may then be evaluated as
\begin{align}
	\chi\left(\omega\right) = \chi_0\sum_{\mathrm{exc}}\frac{\left|\psi_{\mathrm{exc}}\left(0\right)\right|^2}{E_{\mathrm{exc}}\left[E^2_{\mathrm{exc}}-(\hbar\omega+i\hbar\Gamma)^2\right]}
\end{align}
where the sum is taken over all exciton states. Here, $\chi_0$ is a material dependent constant, $\hbar\omega$ is the photon energy, and $\hbar\Gamma$ is a phenomenological line shape broadening. The exciton energy is $E_{\mathrm{exc}} = E_g + E_n$, where $E_g$ is the band gap and $E_n$ the (real) eigenvalues of \cref{eq:Wannierwfield}. This corresponds to measuring the exciton energies from the top of the valence band. For the calculations in the present paper, we use a band gap of $E_g = 2$ eV which is obtained from calculations within the GW approximation in Ref. \cite{Tran2014} and confirmed experimentally by scanning tunneling microscopy/spectroscopy in Ref. \cite{Liang2014}. Additionally, a line broadening of $\hbar\Gamma = 25$ meV is used. The energy eigenvalues are efficiently obtained by expressing the wave function in a finite element basis as described above, and solving the resulting eigenvalue problem. 

The imaginary part of the susceptibility for phosphorene in free space, on an SiO$_2$ substrate, and encapsulated by hBN are shown in the top, middle, and bottom row of \cref{fig:Susceptibility_hBN}, respectively. The structures are subjected to external in-plane electric fields with angles $\alpha\in\left\{0,\frac{\pi}{4},\frac{\pi}{2}\right\}$ to the $x$-axis, indicated in the first, second, and third column, respectively. For each case, field strengths of $0$, $10$, and $20$ V/$\mu$m are considered, and represented by the blue, red and green lines, respectively. As the field strength is increased, the peak absorption corresponding to the fundamental exciton is red-shifted. This is indicated in the zoom plot on the left hand side in each panel. Evidently, the red-shift decreases as the field is rotated from the $x$-axis to the $y$-axis. This is to be expected, as the effective mass in the $y$-direction is larger than in the $x$-direction, and it is therefore more difficult to polarize the fundamental exciton along $y$ than it is along $x$. This is also apparent from the large differences between the $x$- and $y$-components of the polarizabilities shown in \cref{tbl:polarizability}. In addition to the red-shift, the height of the fundamental peaks decreases with increasing field strength. This is natural, as a field pulls electrons and holes in opposite directions, thereby reducing the magnitude of the wave function at the origin. The second peak in the field free spectra corresponds to the third exciton $n=2$. That is, the exciton of type $ce$ (see \cref{tbl:energy}). The reason that the $n=1$ state does not contribute to any peak in the field free spectra is that the state is antisymmetric about the origin (see \cref{eq:state4}), and is therefore zero at the origin. When the field is turned on, these wave functions become polarized and are thus no longer zero at the origin, explaining their contribution to the field-induced absorption spectra. Interestingly, the peak close to the $n=1$ transition energy is more pronounced for an electric field along $y$ than it is for a field along $x$. This is easy to explain considering that the $y$-component of the polarizability is more than twice as large as the $x$-component for this state. It is therefore much easier to polarize along $y$ than it is along $x$. As a final note, we note that the characteristic field induced oscillations above the band gap \cite{Pedersen2002} are clearly visible. Additionally, the field free fundamental peak on an SiO$_2$ substrate at around $1.54$ eV corresponds well with the experimentally observed peak at $1.45$ eV \cite{Liu2014}.

\section{Conclusion}\label{sec:conc}

In the present paper, excitons in phosphorene subjected to an external in-plane electric field have been studied. In particular, we have calculated the unperturbed energies and exciton polarizabilities of the five lowest excitonic states in phosphorene in three different dielectric surroundings. Furthermore, exciton Stark shifts, dissociation rates, and electroabsorption have been computed for various field strengths and directions. A pronounced dependence on the field direction is found for all three quantities. For the fundamental exciton, a field along the armchair axis leads to much more pronounced effects than one along the zigzag direction. For example, the field induced exciton dissociation rates in phosphorene encapsulated by hBN decrease by several orders of magnitude upon rotating the electric field from the armchair to the zigzag axis. This is due to the much larger effective masses found for the zigzag direction than for the armchair direction. An analytical weak-field approximation for the dissociation rate has been derived and shown to agree with the numerically exact rates for weak fields. The larger shift for fields pointing along the armchair direction is again seen by the shift of the fundamental absorption peak when we compute the electroabsorption. On the other hand, the different symmetries of the excited states often lead to larger effects for fields pointing along the zigzag axis, as is apparent from the polarizability tensors.

\begin{acknowledgments}
H.C.K and T.G.P gratefully acknowledge financial support by the Center for Nanostructured Graphene (CNG), which is sponsored by the Danish National Research Foundation, Project No. DNRF103. Additionally, T.G.P. is supported by the QUSCOPE Center, sponsored by the Villum Foundation. N.M.R.P.  acknowledges  support  from  the  European Commission through the project “Graphene-Driven Revolutions in ICT and Beyond” (Reference No. 785219) and the Portuguese Foundation for Science and Technology (FCT) in the framework of the Strategic Financing UID/FIS/04650/2019. In addition, N.M.R.P. acknowledges COMPETE2020, PORTUGAL2020, FEDER, and the FCT through Projects No. PTDC/FIS-NAN/3668/2013, No. POCI-01-0145-FEDER-028114, No. POCI-01-0145-FEDER-029265, No. PTDC/NAN-OPT/29265/2017, and No. POCI-01-0145-FEDER-02888. 
\end{acknowledgments}

\appendix
\section{ANISOTROPIC POISSON EQUATION}\label{app:poisson}

In this appendix, we derive an expression for the interaction energy between two particles in an anisotropic 2D semiconductor. This problem was considered in Ref. \cite{Galiautdinov2019}, where the aim was to obtain closed form expressions for the case of weak anisotropy. In this appendix, we shall not assume weak anisotropy but rather try to obtain as simple an expression for the anisotropic interaction as possible. Note that parts of the derivation are very similar to those in Ref. \cite{Galiautdinov2019} but are included for completeness. The Poisson equation for the potential energy function $V$ between charges $Q$ and $Q'$ located at $(x,y,z)$ and $(0,0,z')$, respectively, may be written as
\begin{align}
\nabla\cdot\left[\boldsymbol{\varepsilon}\cdot\nabla V\left(x,y,z,z'\right)\right] = -4\pi QQ'\delta\left(x\right)\delta\left(y\right)\delta\left(z-z'\right)\thinspace.
\end{align}
where $\boldsymbol{\varepsilon}$ is a dielectric tensor. Assuming the tensor is diagonal and Fourier decomposing the potential energy function as
\begin{multline}
V\left(x,y,z,z'\right) =\\ \frac{1}{4\pi^2}\int_{-\infty}^\infty\int_{-\infty}^\infty\varphi\left(z,z';q_x,q_y\right)e^{i\left(q_xx+q_yy\right)}dq_xdq_y\label{eq:Fourier}
\end{multline}
leads to 
\begin{multline}
\left[\varepsilon_{xx}\left(z\right)q_x^2 +\varepsilon_{yy}\left(z\right)q_y^2-\frac{\partial}{\partial z}\varepsilon_{zz}\left(z\right)\frac{\partial}{\partial z}\right]\varphi\left(z,z';q_x,q_y\right) \\= 4\pi QQ'\delta\left(z-z'\right)\thinspace.
\end{multline}
We take the dielectric functions to be piecewise constant
\begin{align}
\varepsilon_{ii}\left(z\right) = \begin{cases}
\varepsilon_{ii}^{\left(a\right)}\thinspace, \quad z>d/2\\
\varepsilon_{ii}\thinspace, \quad d/2>z>-d/2\thinspace.\\
\varepsilon_{ii}^{\left(b\right)}\thinspace, \quad z<-d/2
\end{cases}
\end{align}
where $ii = \left\{xx,yy,zz\right\}$. Thus, we model the encapsulated 2D sheet as a slab of thickness $d$ surrounded by dielectric media extending to infinity. The solution may then be sought on the form
\begin{align}
\varphi\left(z,z';q_x,q_y\right) = \frac{2\pi QQ'}{q}\begin{cases}
A_1e^{-q_az}\\
A_2e^{-qz}+B_2e^{qz}+\varepsilon_{zz}^{-1}e^{-q\left|z-z'\right|}\\
B_3e^{q_bz}\thinspace,
\end{cases}
\end{align}
where
\begin{align}
q_a &= \left(\frac{\varepsilon_{xx}^{\left(a\right)}q_x^2+\varepsilon_{yy}^{\left(a\right)}q_y^2}{\varepsilon_{zz}^{\left(a\right)}}\right)^{1/2}\thinspace,\\
q &= \left(\frac{\varepsilon_{xx}q_x^2+\varepsilon_{yy}q_y^2}{\varepsilon_{zz}}\right)^{1/2}\thinspace, \\
q_b &= \left(\frac{\varepsilon_{xx}^{\left(b\right)}q_x^2+\varepsilon_{yy}^{\left(b\right)}q_y^2}{\varepsilon_{zz}^{\left(b\right)}}\right)^{1/2}\thinspace.
\end{align}
The Fourier components satisfy the boundary conditions
\begin{align}
\varphi\left(\pm\frac{d}{2},z';q_x,q_y\right) &= \varphi\left(\pm\frac{d}{2},z';q_x,q_y\right)\\
\varepsilon_{zz}^{\left(j\right)}\frac{\partial}{\partial z}\varphi\left(z,z';q_x,q_y\right)|_{z=\pm\frac{d}{2}}&=\varepsilon_{zz}\frac{\partial}{\partial z}\varphi\left(z,z';q_x,q_y\right)|_{z=\pm\frac{d}{2}}\thinspace,
\end{align}
with $j=a$ and $j=b$ for $z=d/2$ and $z=-d/2$, respectively. Enforcing these boundary conditions, placing both charges in the middle of the sheet, and switching to polar coordinates, we obtain
\begin{align}
\varphi\left(0,0;q,\phi\right) = \frac{\varphi_0\left(q\right)}{\varepsilon_{\mathrm{eff}}\left(q,\phi\right)}\thinspace,
\end{align}
where $\varphi_0 = 2\pi QQ'/q$ is the bare interaction and $\varepsilon_{\mathrm{eff}}$ the effective dielectric function given by
\begin{multline}
\varepsilon_{\mathrm{eff}}\left(q,\phi\right) =\\ \frac{g^2\left(g_a+g_b\right)\cosh\left(\frac{dq}{\varepsilon_{zz}}g\right)+g\left(g^2+g_ag_b\right)\sinh\left(\frac{dq}{\varepsilon_{zz}}g\right)}{g^2-g_ag_b+\left(g^2+g_ag_b\right)\cosh\left(\frac{dq}{\varepsilon_{zz}}g\right)+g\left(g_a+g_b\right)\sinh\left(\frac{dq}{\varepsilon_{zz}}g\right)}\thinspace,\label{eq:dielectric_function}
\end{multline}
with
\begin{align}
g_a\left(\phi\right) = \sqrt{\varepsilon_{zz}^{\left(a\right)}\left(\varepsilon_{xx}^{\left(a\right)}\cos^2\phi + \varepsilon_{yy}^{\left(a\right)}\sin^2\phi\right)}\thinspace,\label{eq:g_function}
\end{align}
and $g$ and $g_b$ defined analogously. We now specialize to the case, where the super- and substrate have isotropic in-plane dielectric constants, i.e. $\varepsilon_{xx}^{\left(a\right)} = \varepsilon_{yy}^{\left(a\right)}$ and $\varepsilon_{xx}^{\left(b\right)} = \varepsilon_{yy}^{\left(b\right)}$. This leads to
\begin{align}
g_a = \sqrt{\varepsilon_{zz}^{\left(a\right)} \varepsilon_{xx}^{\left(a\right)} }\thinspace, \qquad	g_b = \sqrt{\varepsilon_{zz}^{\left(b\right)} \varepsilon_{xx}^{\left(b\right)} }\thinspace,
\end{align}
and
\begin{align}
g\left(\phi\right) = \sqrt{\varepsilon_{zz}\left(\varepsilon_{xx}\cos^2\phi + \varepsilon_{yy}\sin^2\phi\right)}\thinspace.
\end{align}
The resulting dielectric function agrees with the one in Ref. \cite{Galiautdinov2019}. Note that the interaction only depends on the dielectric constants of the surrounding media via their geometrical mean. Expanding to first order in $q$, we obtain
\begin{align}
\varepsilon_{\mathrm{eff}}^{\left(1\right)}\left(q,\phi\right) = \kappa + r_0\left(\phi\right)q + O\left(q^2\right)\thinspace.
\end{align}
with
\begin{align}
&\kappa = \frac{\sqrt{\varepsilon_{xx}^{\left(a\right)}\varepsilon_{zz}^{\left(a\right)}} + \sqrt{\varepsilon_{xx}^{\left(b\right)}\varepsilon_{zz}^{\left(b\right)}}}{2}\\ r_0\left(\phi\right) &= \frac{d\left[2\varepsilon_{zz}\left(\varepsilon_{xx}\cos^2\phi + \varepsilon_{yy}\sin^2\phi\right)-\varepsilon_{xx}^{\left(a\right)}\varepsilon_{zz}^{\left(a\right)}-\varepsilon_{xx}^{\left(b\right)}\varepsilon_{zz}^{\left(b\right)}\right]}{4\varepsilon_{zz}}\thinspace.
\end{align}
The first order approximation to the interaction is then
\begin{align}
V\left(q,\phi\right) = \frac{QQ'}{2\pi}\int_{0}^{2\pi}\int_{0}^\infty \frac{e^{iqr\cos\left(\theta-\phi\right)}}{\kappa + r_0\left(\phi\right)q}dqd\phi\thinspace.
\end{align}
The first order dielectric function may be rewritten as
\begin{align}
\varepsilon_{\mathrm{eff}}^{\left(1\right)}\left(q,\phi\right) = \alpha\left(q\right)\left[1-\gamma^2\left(q\right)\cos^2\phi\right]\thinspace,
\end{align}
where
\begin{align}
\alpha\left(q\right) = \kappa+\left( a+r_{0y}\right)q\\
\gamma^2\left(q\right) = \frac{\left(r_{0y}-r_{0x}\right)q}{\alpha\left(q\right)}
\end{align}
with
\begin{align}
a = -\frac{d\left(\varepsilon_a^2+\varepsilon_b^2\right)}{4\varepsilon_{zz}}\thinspace, \quad r_{0x} = \frac{d\varepsilon_{xx}}{2}\thinspace, \quad \text{and} \quad r_{0y} = \frac{d\varepsilon_{yy}}{2}\thinspace.
\end{align}
Here, we have defined screening lengths $r_{0x/y}$ equivalent to the macroscopic definitions in Ref. \cite{Berkelbach2013}. The Fourier series for $1/\varepsilon_{\mathrm{eff}}^{\left(1\right)}$ may be found by the method in Ref. \cite{Mikhlin1964}. We get 
\begin{align}
\frac{1}{\varepsilon_{\mathrm{eff}}^{\left(1\right)}} = \frac{1}{\alpha\sqrt{1-\gamma^2}}\left[1+2\sum_{k=1}^{\infty}\left(\frac{\gamma}{1+\sqrt{1-\gamma^2}}\right)^{2k}\cos\left(2k\phi\right)\right]\thinspace.
\end{align}
The angular integral in the interaction may then be written as
\begin{multline}
I\left(q\right) = \int_{0}^{2\pi}\frac{e^{iqr\cos\left(\phi-\theta\right)}}{\varepsilon_{\mathrm{eff}}^{\left(1\right)}\left(q,\phi\right)}d\phi\\
= 	\int_{0}^{2\pi}\frac{e^{iqr\cos\left(\phi-\theta\right)}}{\alpha\sqrt{1-\gamma^2}}\left[1+2\sum_{k=1}^{\infty}\left(\frac{\gamma}{1+\sqrt{1-\gamma^2}}\right)^{2k}\cos\left(2k\phi\right)\right]\thinspace,
\end{multline}
which leads to
\begin{multline}
I\left(q\right) = \frac{2\pi}{\alpha\sqrt{1-\gamma^2}}\left.\biggl[J_0\left(qr\right)\right.\\\left.+2\sum_{k=1}^{\infty}\left(-1\right)^k\left(\frac{\gamma}{1+\sqrt{1-\gamma^2}}\right)^{2k}J_{2k}\left(qr\right)\cos\left(2k\theta\right)\biggr]\right.\thinspace.
\end{multline}
Now, $a$ is typically very small compared to $r_{0y}$, i.e. $a\ll r_{0y}$, and we get $\alpha\approx\kappa+r_{0y}q$. Defining
\begin{align}
\varepsilon_x = \kappa + r_{0x}q \quad \text{and}\quad 	\varepsilon_y = \kappa + r_{0y}q\thinspace
\end{align}
we may write
\begin{align}
\frac{1}{\varepsilon_{\mathrm{eff}}^{\left(1\right)}}\approx \frac{1}{\sqrt{\varepsilon_x\varepsilon_y}}\left[1+2\sum_{k=1}^{\infty}\left(\frac{\sqrt{\varepsilon_y}-\sqrt{\varepsilon_x}}{\sqrt{\varepsilon_y}+\sqrt{\varepsilon_x}}\right)^k\cos\left(2k\phi\right)\right]\thinspace
\end{align}
and the integral as
\begin{multline}
I\left(q\right) \approx\\ \frac{2\pi}{\sqrt{\varepsilon_x\varepsilon_y}}\left\{J_0\left(qr\right)+2\sum_{k=1}^{\infty}\left(\frac{\sqrt{\varepsilon_x}-\sqrt{\varepsilon_y}}{\sqrt{\varepsilon_x}+\sqrt{\varepsilon_y}}\right)^kJ_{2k}\left(qr\right)\cos\left(2k\theta\right)\right\}\thinspace.
\end{multline}
The full interaction with a linearized dielectric function may thus be computed as
\begin{align}
V\left(r,\theta\right) = \frac{QQ'}{2\pi}\int_{0}^\infty I\left(q\right)dq\thinspace.
\end{align}
\section{ANISOTROPIC WEAK-FIELD ASYMPTOTIC THEORY}\label{app:WFAT}

In this appendix, we show how the weak-field asymptotic theory (WFAT) of tunneling ionization \cite{Tolstikhin2011,Batishchev2010} may be extended to a two-dimensional system with an anisotropic potential. The aim of this appendix is thus to derive an analytical weak-field approximation for the ionization rate of the auxiliary system defined by 
\begin{align}
\left[-\frac{1}{2}\nabla^2 + V\left(x,y\right) 
+ \mathcal{E}x - E\right]\psi\left(x,y\right)=0\thinspace,\label{eq:aux}
\end{align}
where we assume that 
\begin{align}
	\lim\limits_{x\to-\infty}-rV\left(x,y\right) = Z_{\mathrm{asymp}}\thinspace,\label{eq:assumption}
\end{align}
with $Z_{\mathrm{asymp}}>0$ a real constant. Note that $V$ need not be isotropic for this condition to be satisfied. In parabolic cylindrical coordinates
\begin{align}
&u = r+x\thinspace,\quad u\in\left[0,\infty\right)\\
&v=r-x\thinspace,\quad v\in\left[0,\infty\right)\thinspace,\label{eq:parav}
\end{align}
\cref{eq:aux} reads
\begin{align}
\left[ \sqrt{v}\frac{\partial}{\partial v}\sqrt{v}\frac{\partial }{\partial v} +  \frac{\mathcal{E} v^2}{4} +\frac{Ev}{2} + \beta\left(v\right)\right]\psi\left(u,v\right)=0\thinspace,\label{eq:paraaux}
\end{align}
where
\begin{align}
\beta\left(v\right)=\sqrt{u}\frac{\partial}{\partial u}\sqrt{u}\frac{\partial }{\partial u} -rV\left(x,y\right)  +\frac{Eu}{2}-\frac{\mathcal{E} u^2}{4}
\end{align}
operates on functions of $u$ and depends on $v$ as a parameter through $V$. It has a purely discrete spectrum defined by
\begin{align}
\beta\left(v\right)\varphi_n\left(u;v\right) = b_n\varphi_n\left(u;v\right)\thinspace.
\end{align}
It is symmetric (but not hermitian due to generally complex $E$) with respect to the weighting function $w\left(u\right) = 1/\sqrt{u}$, and we may therefore choose the eigenfunctions orthonormal
\begin{align}
\left(\varphi_n|\varphi_m\right)_{u,w} = \int_{0}^{\infty}\varphi_n\left(u;v\right) \varphi_m\left(u;v\right)\frac{1}{\sqrt{u}}du = \delta_{nm}\thinspace,
\end{align}
where we have used regular parentheses for the inner product to indicate that there is no complex conjugation, which is a general property of the theory of Siegert states \cite{Siegert1936,Tolstikhin1998,Sitnikov2003,Toyota2005,Batishchev2007}. The subscript $u,w$ denotes that the integral is taken with respect to $u$, using the weighting function $w$. 

We shall proceed by writing the solution to \cref{eq:paraaux} as
\begin{align}
\psi\left(\boldsymbol{r}\right) = \sum_n v^{-1/4}f_n\left(v\right)\varphi_n\left(u;v\right)\thinspace.\label{sir:eq:expansion}
\end{align}
This approach is based on the adiabatic expansion applied to a three-dimensional system in Refs. \cite{Batishchev2010,Tolstikhin2011}. In essence, it corresponds to treating $v$ as a slow variable, much like the internuclear distance in the Born-Oppenheimer approximation. It should be noted, however, that the expansion does not constitute an approximation as long as all nonadiabatic coupling terms are taken into account. Substituing the expansion into \cref{eq:paraaux}, we obtain
\begin{multline}
\left[ \frac{\partial^2}{\partial v^2} +  \frac{\mathcal{E} v}{4} +\frac{E}{2} + \frac{b_n\left(v\right)}{v}+\frac{3}{16v^2}\right]f_n\left(v\right) \\+ \sum_n \left[\left(\varphi_m\left|\frac{\partial \varphi_n }{\partial v}\right.\right)_{u,w}2\frac{\partial }{\partial v} +\left(\varphi_m\left|\frac{\partial^2 \varphi_n }{\partial v^2}\right.\right)_{u,w}\right]f_n\left(v\right)=0\thinspace.
\end{multline}
Recalling the assumption in \cref{eq:assumption} [and noting that $x = (u-v)/2$ ], the $\varphi$ functions cease to depend on $v$ for $v\to\infty$ and the coupling matrix elements therefore reduce to zero. Explicitly
\begin{align}
\left[ \frac{\partial^2}{\partial v^2} +  \frac{\mathcal{E} v}{4} +\frac{E}{2} + \frac{b_n}{v}+\frac{3}{16v^2}\right]f_n\left(v\right) = 0\thinspace.\label{eq:fdiff}
\end{align}
This equation is identical to the one in Refs. \cite{Tolstikhin2011} and \cite{Trinh2013}. For $\mathcal{E}=0$ the solutions behaves as 
\begin{align}
f^{\left(0\right)}\left(v\right) = v^{b_n/k}e^{-kv/2}\left[1 + \frac{c_1}{v}+\frac{c_2}{v^2}+O\left(v^{-3}\right)\right]\thinspace.\label{sir:eq:h0}
\end{align}
For $\mathcal{E}>0$, the outgoing solution satisfies \cite{Batishchev2010}
\begin{align}
&f_n\left(v\right)_{v\rightarrow \infty} = c_n f\left(v\right)\label{eq:fasymp}\\
f\left(v\right) &= \frac{\sqrt{2}}{\left(\mathcal{E} v\right)^{1/4}}\exp\left[\frac{i\sqrt{\mathcal{E}}v^{3/2}}{3}+\frac{iE\sqrt{v}}{\sqrt{\mathcal{E}}}\right]\thinspace.
\end{align}
Thus, the asymptotics only depend on $b_n$ through the coefficients $c_n$. 
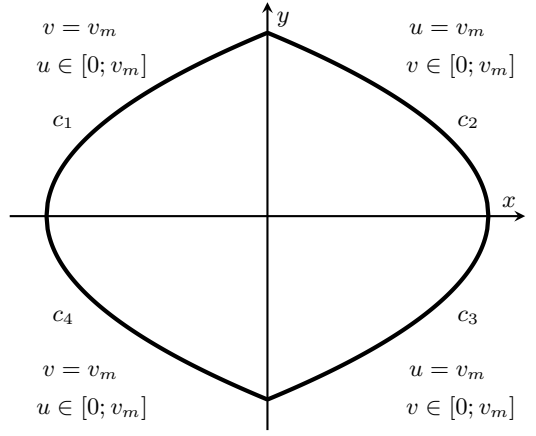
\begin{figure}
	\centering
	\begin{tikzpicture}[scale=1]
	%\draw [->,thick] (0,-1) -- (0,1) node (yaxis) [below left] {$y$};
	%\draw [->, thick] (-1.4,0) -- (1.4,0) node (xaxis) [below right] {$x$};
	%\draw[scale=0.5,domain=-1:1,smooth,variable=\x,blue] plot ({\x},{\x*\x});
	\begin{axis}[
	xmin=-3.5,xmax=3.5,
	ymin=-3.5,ymax=3.5,
	xtick=\empty,
	ytick=\empty,
	axis lines=center,
	thick,
	xlabel=$x$,
	ylabel=$y$,
	]
	\node at (axis cs:-2.5,1.8) [anchor=north east] {$c_1$};
	\node at (axis cs:-1.9,3.3) [anchor=north east] {$v=v_m$};
	\node at (axis cs:-1.5,2.8) [anchor=north east] {$u\in\left[0;v_m\right]$};
	\addplot [domain=0:3,samples=200,ultra thick]({x-3},{sqrt(x*3)}); 
	
	\node at (axis cs:3,1.8) [anchor=north east] {$c_2$};
	\node at (axis cs:3.1,3.3) [anchor=north east] {$u=v_m$};
	\node at (axis cs:3.5,2.8) [anchor=north east] {$v\in\left[0;v_m\right]$};
	\addplot [domain=0:3,samples=200,ultra thick]({3-x},{sqrt(x*3)});

	\node at (axis cs:3,-1.4) [anchor=north east] {$c_3$};
	\node at (axis cs:3.1,-2.3) [anchor=north east] {$u=v_m$};
	\node at (axis cs:3.5,-2.8) [anchor=north east] {$v\in\left[0;v_m\right]$};
	\addplot [domain=0:3,samples=200,ultra thick]({3-x},{-sqrt(x*3)});
	
	\node at (axis cs:-2.5,-1.4) [anchor=north east] {$c_4$};
	\node at (axis cs:-1.9,-2.3) [anchor=north east] {$v=v_m$};
	\node at (axis cs:-1.5,-2.8) [anchor=north east] {$u\in\left[0;v_m\right]$};
	\addplot [domain=0:3,samples=200,ultra thick]({x-3},{-sqrt(x*3)});

	\end{axis}
	
	%\draw [red, thick,  domain=--1:1, samples=40] 
	%plot ({\x-1}, {\x} );
	%\draw [ultra thick] (0.8,-0.5) -- (-0.8,-0.5);
	%\draw [ultra thick] (0.8,0.5) -- (-0.8,0.5);
	%\draw [ultra thick] (-.8,-0.5) -- (-0.8,0.5);
	%\draw [ultra thick] (.8,-0.5) -- (0.8,0.5);
	%\node at (1.2,0.3) [below] {$V = \infty$};
	%\node at (0,-0.3) [below] {$V = 0$};
	% Draw two intersecting lines
	% Draw axes
	%|- (3,0) node (xaxis) [right] {$z$};
	% Calculate the intersection of the lines a_1 -- a_2 and b_1 -- b_2
	% and store the coordinate in c.
	% Draw lines indicating intersection with y and x axis. Here we use
	% the perpendicular coordinate system
	% Draw a dot to indicate intersection point
	\end{tikzpicture}
	\caption{The area enclosed by curves of constant $u$ and $v$ respectively. Note that the curves on the lower half-plane correspond to the negative sign in $y$.}\label{sir:fig:area}
\end{figure}
The dissociation rate may be related to the probability current $\boldsymbol{j}$ as follows
\begin{align}
\Gamma \left|\psi\right|^2 = \nabla\cdot\left[-\frac{i}{2}\left(\psi^*\nabla \psi - \psi \nabla \psi^*\right)\right] = \nabla \cdot\boldsymbol{j}\thinspace.\label{sir:eq:gamma}
\end{align}
In the weak field region, the resonance state $\psi$ will coincide with the unperturbed bound state in a region $v<v_m$, where $v_m$  is defined by $v_t\ll v_m\ll\mathcal{E}/\Gamma^2$\thinspace. Here $v_t\approx -2E_0/\mathcal{E}$ (see \cref{eq:fdiff}) is the turning point, and the fact that this holds can be seen from the fact that the exponential growth of $f_n$ starts at around $v \gtrapprox \mathcal{E}/\Gamma^2$. Let $A$ be the area enclosed by curves of constant $u=v_m$ and $v=v_m$ respectively (see \cref{sir:fig:area}). Then inside this area the SS will be approximately equal to the unperturbed state and therefore 
\begin{align}
\int_A\left|\psi\right|^2dA \approx 1\thinspace,
\end{align}
and integrating both sides of \cref{sir:eq:gamma} over the area $A$ therefore yields 
\begin{align}
\Gamma = \int_A\nabla \cdot\boldsymbol{j}dA = \int_c \hat{n}\cdot \boldsymbol{j}dl\thinspace,
\end{align}
where the final equality follows from the divergence theorem. Now, if $v_m$ is large enough, integrating along $c_2$ and $c_3$ will yield zero because $\varphi_n \rightarrow 0$ for $u \rightarrow \infty$. The integration therefore immediately reduces to integrating along $c_1$ and $c_4$ (see \cref{sir:fig:area}). By symmetry, the integration along $c_4$ must equal the integration along $c_1$, and thus we get 
\begin{align}
\Gamma =  2\int_{c_1} \hat{v}\cdot \boldsymbol{j}dc_1\thinspace.
\end{align}
The parametrization of the curve $c_1$ is given by
\begin{align}
\boldsymbol{r} = \frac{u-v_m}{2}\hat{x} + \sqrt{uv_m}\hat{y}\thinspace,
\end{align}
and
\begin{align}
\left|\frac{\partial}{\partial u}\boldsymbol{r}\right| = \sqrt{\frac{r}{2u}}\thinspace.
\end{align}
The normal vector $\hat{v}$ picks out the $v$ component of the gradient in $\boldsymbol{j}$ so that
\begin{align}
\hat{v}\cdot\boldsymbol{\nabla} = \sqrt{\frac{2v}{r}}\frac{\partial}{\partial v}\thinspace,
\end{align}
and we have
\begin{multline}
\Gamma = -i\int_{0}^{v_m}\sqrt{\frac{v_m}{u}}\left[\psi^*\left(u,v_m\right)\frac{\partial}{\partial v} \psi\left(u,v_m\right) \right. \\  \left.- \psi\left(u,v_m\right) \frac{\partial}{\partial v} \psi^*\left(u,v_m\right)\right] du\thinspace,
\end{multline}
where $\partial \psi\left(u,v_m\right)/\partial v$ denotes the derivative of $\psi$ evaluated at the point $v_m$. For sufficiently weak fields, $v_m$ will become so large that we might take the limit $v_m\rightarrow\infty$. This allows us to use the asymptotic basis functions that are independent of $v$ in the expression. We obtain
\begin{align}
\Gamma = -i\sum_{n}\left[ f_n^*\left(v_m\right)\frac{\partial}{\partial v}  f_n\left(v_m\right)- f_n\left(v_m\right) \frac{\partial}{\partial v}  f_n^*\left(v_m\right)\right] \thinspace.
\end{align}
Recalling the asymptotic expression for $f_n$ in \cref{eq:fasymp}, one obtains
\begin{align}
\Gamma = 2\sum_n \left|c_n\right|\left(1+\frac{E_R}{\mathcal{E} v_m}\right)\exp\left(\Gamma \sqrt{\frac{v_m}{\mathcal{E}}}\right)\thinspace.
\end{align}
In the weak field region, $\Gamma$ will be exponentially small and we can approximate the exponential function by unity. Further more for $v_m$ large enough the second term in the parenthesis can be neglected (recall one of the assumptions we made was $v_m\gg v_t\approx-2E_0/\mathcal{E}$). Ultimately this leads to
\begin{align}
\Gamma = 2\sum_n \left|c_n\right| \thinspace.\label{eq:disssum}
\end{align}
Thus, to find the weak field ionization rate, all we require is the coefficients $c_n$. We shall find those by matching the state with a field present to the unperturbed state in a matching region that is far enough from the origin that the asymptotics apply, yet close enough that the unperturbed and the perturbed states coincide. We may write the unperturbed state as
\begin{align}
\psi_0\left(\boldsymbol{r}\right) = \sum_n g_nv^{-1/4}f_n^{\left(0\right)}\left(v\right)\varphi_n^{\left(0\right)}\left(u\right)\quad \text{for } v\to \infty \thinspace,
\end{align}
where $g_n$ is a field independent coefficient. Recalling the solution for $\mathcal{E}=0$ is given by \cref{sir:eq:h0}, we have
\begin{align}
\psi_0\left(\boldsymbol{r}\right) = \sum_n g_n v^{b_n/k-1/4}e^{-kv/2}\varphi_n^{\left(0\right)}\left(u\right) \quad \text{for } v\to\infty\thinspace.\label{sir:eq:unperturbed}
\end{align}
In the matching region, a WKB type expression for the perturbed $f_n$ functions can be found to be \cite{Tolstikhin2011}
\begin{multline}
f_n\left(v\right) = c_n\sqrt{\frac{2}{k}}\left(\frac{\mathcal{E}}{4k^2}\right)^{b_n^{\left(0\right)}/k}\\ \times \exp\left[-\frac{i\pi}{4}-\frac{i\pi b_n^{\left(0\right)} }{k} + \frac{k^3}{3\mathcal{E}}\right]v^{b_n/k}e^{-kv/2}\thinspace.\label{sir:eq:WKB}
\end{multline}
Comparing \cref{sir:eq:WKB} to \cref{sir:eq:unperturbed} leads to the conclusion that
\begin{align}
c_n = g_n\sqrt{\frac{k}{2}}\left(\frac{4k^2}{\mathcal{E}}\right)^{b_n^{\left(0\right)}/k}\exp\left[\frac{i\pi}{4}+\frac{i\pi b_n^{\left(0\right)} }{k} - \frac{k^3}{3\mathcal{E}}\right]\thinspace,\label{sir:eq:cn}
\end{align}
and therefore
\begin{align}
\Gamma = \sum_n\left|g_n\right|^2k\left(\frac{4k^2}{\mathcal{E}}\right)^{2b_n^{\left(0\right)}/k}\exp\left(- \frac{2k^3}{3\mathcal{E}}\right)\thinspace.
\end{align}
What remains is to find the field-free eigenvalues $b_n^{\left(0\right)}$. For $\mathcal{E}=0$, we have
\begin{align}
\left(\sqrt{u}\frac{\partial}{\partial u}\sqrt{u}\frac{\partial }{\partial u} +Z_{\mathrm{asymp}} -\frac{k^2u}{4}-b_n^{\left(0\right)}\right)\varphi_n^{\left(0\right)}\left(u\right) = 0\thinspace.
\end{align}
We find
\begin{align}
\varphi_n^{\left(0\right)}\left(u\right) = N_{n} L_{n}^{\left(-1/2\right)}\left(ku\right)e^{-ku/2}\thinspace,
\end{align} 
where  $n=0,1,2\dots$ and the normalization coefficient
\begin{align}
N_n = 	\left[\frac{k^{1/2}n!}{\left(n-1/2\right)!}\right]^{1/2}\thinspace.
\end{align}
The eigenvalues are
\begin{align}
b_n^{\left(0\right)} = Z_{\mathrm{asymp}}-k\left(n+\frac{1}{4}\right)\thinspace, \quad \mathrm{where}\,\, n=0,1,2\dots\thinspace,
\end{align}
and the dissociation rate becomes
\begin{align}
\Gamma = \sum_n\left|g_n\right|^2k\left(\frac{4k^2}{\mathcal{E}}\right)^{2Z_{\mathrm{asymp}}/k-2n-1/2}\exp\left(- \frac{2k^3}{3\mathcal{E}}\right)\thinspace.
\end{align}
As discussed in Ref. \cite{Tolstikhin2011}, only the dominant contribution may be included in \cref{eq:disssum} within the present approximation. It corresponds to $n=0$, thus the weak-field approximation to the dissociation rate is
\begin{align}
\Gamma = \left|g_0\right|^2k\left(\frac{4k^2}{\mathcal{E}}\right)^{2Z_{\mathrm{asymp}}/k-1/2}\exp\left(- \frac{2k^3}{3\mathcal{E}}\right)\thinspace.
\end{align}
The coefficient $g_0$ is defined by the asymptotics of the unperturbed state. It can be obtained by taking the inner product between \cref{sir:eq:unperturbed} and $\varphi_0^{\left(0\right)}$, i.e.
\begin{align}
g_0 = \lim\limits_{v\to\infty}v^{1/2-Z/k}e^{kv/2}\int_{0}^{\infty}\varphi_0^{\left(0\right)}\left(u\right)\psi_0\left(\frac{u+v}{2}\right)\frac{1}{\sqrt{u}}du\thinspace.\label{eq:g0}
\end{align}

\bibliography{litt}

\end{document}